\documentclass[sigconf]{acmart}
\usepackage{array}
\AtBeginDocument{%
  }

\copyrightyear{2025}
\acmYear{2025}
\setcopyright{cc}
\setcctype{by}
\acmConference[PEARC '25]{Practice and Experience in Advanced Research Computing}{July 20--24, 2025}{Columbus, OH, USA}
\acmBooktitle{Practice and Experience in Advanced Research Computing (PEARC '25), July 20--24, 2025, Columbus, OH, USA}
\acmDOI{10.1145/3708035.3736086}
\acmISBN{979-8-4007-1398-9/2025/07}

\begin{document}

\title{Real-Time In-Network Machine Learning on P4-Programmable FPGA SmartNICs with Fixed-Point Arithmetic and Taylor Approximations}

\author{Mohammad Firas Sada}
\email{mfsada@ucsd.edu}
\orcid{0009-0006-6045-2940}
\affiliation{%
  \institution{University of California, San Diego}
  \city{La Jolla}
  \state{CA}
  \country{USA}
}

\author{John J. Graham}
\email{jjgraham@ucsd.edu}
\orcid{0000-0002-2139-5617}
\affiliation{%
  \institution{University of California, San Diego}
  \city{La Jolla}
  \state{CA}
  \country{USA}
}

\author{Mahidhar Tatineni}
\email{mahidhar@sdsc.edu}
\orcid{0009-0003-0709-090X}
\affiliation{%
  \institution{University of California, San Diego}
  \city{La Jolla}
  \state{CA}
  \country{USA}
}

\author{Dmitry Mishin}
\email{dmishin@ucsd.edu}
\orcid{0000-0003-1125-448X}
\affiliation{%
  \institution{University of California, San Diego}
  \city{La Jolla}
  \state{CA}
  \country{USA}
}


\author{Thomas A. DeFanti}
\email{tdefanti@ucsd.edu}
\orcid{0000-0002-7642-2336}
\affiliation{%
  \institution{University of California, San Diego}
  \city{La Jolla}
  \state{CA}
  \country{USA}
}

\author{Frank Würthwein}
\email{fkw@physics.ucsd.edu}
\orcid{0000-0001-5912-6124}
\affiliation{%
  \institution{University of California, San Diego}
  \city{La Jolla}
  \state{CA}
  \country{USA}
}


\renewcommand{\shortauthors}{Sada et al.}

\begin{abstract}
As machine learning (ML) applications become integral to modern network operations, there is an increasing demand for network programmability that enables low-latency ML inference for tasks such as Quality of Service (QoS) prediction and anomaly detection in cybersecurity. ML models provide adaptability through dynamic weight adjustments, making Programming Protocol-independent Packet Processors (P4)-programmable\cite{P4} FPGA SmartNICs an ideal platform for investigating In-Network Machine Learning (INML). These devices offer high-throughput, low-latency packet processing and can be dynamically reconfigured via the control plane, allowing for flexible integration of ML models directly at the network edge. This paper explores the application of the P4 programming paradigm to neural networks and regression models, where weights and biases are stored in control plane table lookups. This approach enables flexible programmability and efficient deployment of retrainable ML models at the network edge, independent of core infrastructure at the switch level.
\end{abstract}


\begin{teaserfigure}
    \centering
    \includegraphics[width=1\textwidth]{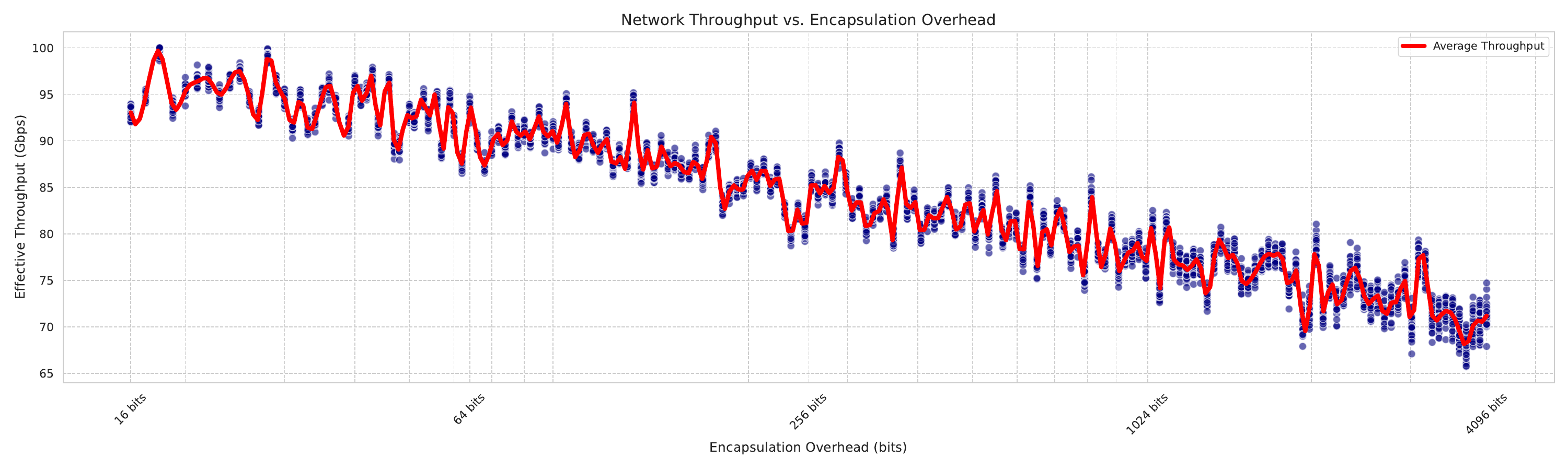}
\caption{Impact of encapsulation header overhead (input features for regression models) on network throughput (Gbps) in a 100Gbps FPGA SmartNIC. Throughput for ingress and egress traffic is measured as a function of header bit size, which corresponds to the inclusion of input features (data embedded in packets for regression models) and directly affects processing efficiency. Output features (model predictions) are generated during egress and returned to the network. The FPGA model, implemented in P4, uses table lookups for retrainable models via the control plane, sacrificing some throughput for flexibility. To reduce arbitration, we assume input features and weights follow the same fractional and integer bits. The x-axis (logarithmic scale) shows encapsulation overhead (bits), and the y-axis shows measured throughput, with scatter points indicating individual measurements and a red line denoting average performance. Increasing overhead reduces throughput due to added processing demands, highlighting the trade-off between input features and network efficiency.}
    \label{fig:third_plot}
\end{teaserfigure}


\maketitle

\section{Introduction}
\label{sec:introduction}
The integration of machine learning (ML) into network infrastructure has emerged as a critical enabler for real-time applications. However, traditional ML deployment models—reliant on centralized GPU/CPU clusters— introduce latency bottlenecks that hinder their effectiveness in time-sensitive, high-throughput environments. P4-Programmable network hardware, such as FPGA P4 SmartNICs, offers a transformative alternative by enabling computation directly within the data plane. Yet, the absence of native floating-point support and limited arithmetic operations in protocol-independent switch and SmartNIC architecture (P4) poses significant barriers to deploying ML models at line rate. This paper bridges this gap by presenting a novel framework for in-network ML inference that uses mathematical methods to convert ML models into programmable data planes and control planes deployed on the National Research Platform’s FPGA SmartNIC infrastructure.

\subsection{The National Research Platform (NRP)}
\label{sec:nrp}
    The National Research Platform (NRP) is a distributed, multi-tenant, Kubernetes-based cyberinfrastructure designed to facilitate collaborative scientific computing. Spanning over 75 sites internationally, the NRP uniquely integrates diverse computational resources, ranging from single nodes to extensive GPU and CPU clusters, to support various research workloads, including advanced AI and machine learning tasks. It emphasizes flexibility through user-friendly interfaces such as JupyterHub. The NRP includes more than 1,400 GPUs, utilized by over 300 research groups and numerous classrooms. More importantly, its 32 AMD/Xilinx Alveo U55C FPGAs provide immense capability for programmable network applications \cite{smarr_pacific_2018}. Each FPGA, equipped with two fully P4-programmable 100Gbps ports (using the AMD OpenNIC shell) and two PCIe-connected host ports, functions as a SmartNIC.

    \subsection{Motivation}
    \label{subsec:motivation}
The growing need for real-time, low-latency ML inference in network environments, such as academic clusters, drives the demand for decentralized, edge-centric intelligence. Traditional approaches that offload ML computations to CPUs or GPUs via PCIe interfaces introduce significant latency, undermining their suitability for time-sensitive tasks like intrusion detection or QoS optimization. While FPGA SmartNICs are often used as PCIe-based accelerators, their true potential lies in enabling in-network computation through P4 programmability. However, deploying ML models directly on the data plane faces challenges, including P4’s lack of native support for floating-point arithmetic and regression operations. Our work addresses these limitations by approximating floating-point operations through fixed-point arithmetic using Taylor series expansions\cite{patel}, enabling ML inference within the constraints of P4’s syntax and hardware. By embedding models into the data plane, we reduce PCIe communication from overheads by relying on control plane table lookups solely for the weights/coefficients, achieving microsecond-scale inference times critical for high-performance networks. This methodology not only enhances programmability but also opens avenues for adaptive cybersecurity frameworks and QoS mechanisms that evolve with dynamic network conditions, marking a significant step toward self-optimizing network infrastructures.

\section{Methodology}

The proposed approach uses the Alveo U55C FPGAs within the NRP for in-network inference. The ESnet SmartNIC tool stack~\cite{esnet_cite} is used to provide an integrated software environment for OpenNIC shell, incorporating features such as DPDK-pktgen~\cite{dpdk_cite}, probe counters, and automated FPGA flashing, all managed through a Docker Compose configuration~\cite{docker_cite}. 

The methodology consists of a software pipeline that transforms trained Python-based regression models into a format suitable for execution within the P4 data plane. After training, model weights and biases are serialized into a structured textual format, which is then parsed to generate P4 control plane table entries. These entries store the model parameters, including weight scaling factors, in a manner compatible with conventional P4 targets such as BMv2 \cite{paolini2021programmable}. The entire P4 syntax and control plane configuration are automatically synthesized and deployed onto the FPGA.

First, accuracy validation and computational trade-offs are assessed through software simulations. Initially, the trained model is executed on a CPU using Python to evaluate fixed-point arithmetic approximations and loss characteristics. Following this, BMv2 simulations~\cite{bmv2_cite} are conducted, utilizing traffic generated via Scapy~\cite{scapy_cite}, to verify correctness and expected packet behavior.

FPGA inference operates within the data plane, processing network packets with an appended header. Input features are extracted, and model weights are retrieved from control plane tables, eliminating registers/externs. After computation, the header is replaced with an output format for interoperability \cite{sapio2023janus}.

For low-latency ML inference in academic clusters like NRP\cite{smarr_pacific_2018}, the motivation is to execute neural network inference directly in the P4 data plane. We approximate floating-point operations using fixed-point arithmetic and Taylor series expansions, avoiding PCIe offloading.

\begin{figure}[ht]
    \centering
    \includegraphics[width=0.7\linewidth]{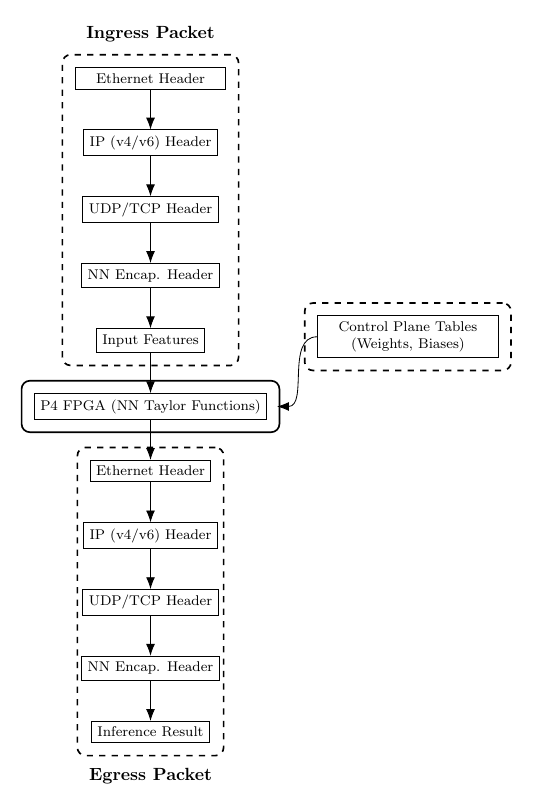}
    \caption{Packet processing pipeline for neural network inference using P4-programmable FPGAs. The control plane loads model parameters; packets carry input features, are processed via Taylor-approximated functions, and exit with embedded inference results.}
    \label{fig:packet_processing}
\end{figure}

\begin{table}[ht]
    \centering
    \caption{Neural Network Encapsulation Header}
    \label{tab:nn_header}
    \renewcommand{\arraystretch}{0.6}
    \footnotesize
    \begin{tabular}{>{\centering\arraybackslash}m{1.4cm} | >{\centering\arraybackslash}m{1.2cm} | >{\centering\arraybackslash}m{3cm}}
        \toprule
        \textbf{Field} & \textbf{Size (bits)} & \textbf{Description} \\
        \midrule
        Model ID & 16 & Model identifier \\
        Feature Cnt & 8 & \# input features \\
        Output Cnt & 8 & \# output features \\
        Scale & 16 & Fixed-point scaling factor \\
        Flags & 8 & Control flags (e.g., padding) \\
        \midrule
        Feature 1 & 32 & 1st input feature value \\
        Feature 2 & 32 & 2nd input feature value \\
        $\vdots$ & $\vdots$ & $\vdots$ \\
        Feature N & 32 & $N$th input feature value \\
        \bottomrule
    \end{tabular}
\end{table}

\section{Mathematical Details}
This section details the mathematical foundations of the model.
\begin{enumerate}
    \item \textbf{Floating-Point to Fixed-Point Conversion:}
    \begin{itemize}
        \item Floating-point $x$ to fixed-point $x_q$ via scaling $2^s$. Equations and parameters are in Table~\ref{tab:fixed_point_complex}.
    \end{itemize}
    \item \textbf{Taylor Series Approximations:}
    \begin{itemize}
        \item Non-linear functions (e.g., sigmoid) approximated using Taylor series. Table~\ref{tab:taylor_sigmoid_complex} compares orders, and Table~\ref{tab:constants_complex} lists scaled constants.
    \end{itemize}
    \item \textbf{Control Plane Integration:}
    \begin{itemize}
        \item Fixed-point parameters and Taylor coefficients stored in FPGA control plane tables (Table~\ref{tab:constants_complex}) for efficient retrieval.
    \end{itemize}
\end{enumerate}

\subsection{Fixed-Point Representation}

Table~\ref{tab:fixed_point_complex} summarizes the procedures for encoding a floating-point weight into fixed-point and decoding it back. Here, a weight $w$ is encoded as follows in the table.

\begin{table}[ht]
    \centering
    \caption{Fixed-Point Encoding and Decoding}
    \label{tab:fixed_point_complex}
    \renewcommand{\arraystretch}{0.7}
    \footnotesize
    \begin{tabular}{>{\centering\arraybackslash}m{1.2cm} | >{\centering\arraybackslash}m{2.8cm} | >{\centering\arraybackslash}m{3.1cm}}
        \toprule
        \textbf{Operation} & \textbf{Equation} & \textbf{Parameters} \\
        \midrule
        Encoding & $w_q = \text{round}(w \times 2^s) + b$ & $w$: float weight,\newline $s$: scale,\newline $b$: offset \\
        Decoding & $w \approx \frac{w_q - b}{2^s}$ & $w_q$: fixed-point,\newline $s$: scale,\newline $b$: offset \\
        \bottomrule
    \end{tabular}
\end{table}

\subsection{Sigmoid Approximation}

The sigmoid activation function is approximated using Taylor series expansions to facilitate its implementation in fixed-point arithmetic. Table~\ref{tab:taylor_sigmoid_complex} presents the first-order (linear), third-order (cubic), and fifth-order (quintic) approximations, where $R_n(x)$ represents the residual error term at each order.

\begin{table}[ht]
    \centering
    \caption{Taylor Series Approximations for Sigmoid with Error Terms}
    \label{tab:taylor_sigmoid_complex}
    \renewcommand{\arraystretch}{0.7}
    \footnotesize
    \begin{tabular}{>{\centering\arraybackslash}m{1.2cm} | >{\arraybackslash}m{3.8cm} | >{\arraybackslash}m{2.8cm}}
        \toprule
        \textbf{Order} & \textbf{Approximation} & \textbf{Use Case} \\
        \midrule
        1st (Linear) & $\sigma(x) \approx 0.5 + \frac{x}{4} + R_1(x)$ & Low-precision for small $|x|$; residual error $R_1(x)$. \\
        3rd (Cubic) & $\sigma(x) \approx 0.5 + \frac{x}{4} - \frac{x^3}{48} + R_3(x)$ & Higher precision over broader range. \\
        5th (Quintic) & $\sigma(x) \approx 0.5 + \frac{x}{4} - \frac{x^3}{48} + \frac{x^5}{1440} + R_5(x)$ & Very high precision over wider inputs. \\
        \bottomrule
    \end{tabular}
\end{table}

Table~\ref{tab:constants_complex} lists the scaled constants used in the fixed-point arithmetic when a typical scaling factor of $s = 16$ is employed.

\begin{table}[ht]
    \centering
    \caption{Scaled Constants for Fixed-Point Arithmetic with Higher Precision}
    \label{tab:constants_complex}
    \renewcommand{\arraystretch}{0.7}
    \footnotesize
    \begin{tabular}{>{\centering\arraybackslash}m{2.6cm} | >{\centering\arraybackslash}m{2.6cm} | >{\centering\arraybackslash}m{2.6cm}}
        \toprule
        \textbf{Constant} & \textbf{Float Value} & \textbf{Fixed-Point ($s = 16$)} \\
        \midrule
        Bias & 0.5 & $32768$ \\
        Linear Term ($\frac{1}{4}$) & 0.25 & $16384$ \\
        Cubic Term ($-\frac{1}{48}$) & $-0.0208333$ & $-1365$ \\
        Quintic Term ($\frac{1}{1440}$) & $0.0006944$ & $45$ \\
        \bottomrule
    \end{tabular}
\end{table}

\subsection{ReLU and Piecewise Linear Approximations}

Activation functions such as ReLU and its variants (Leaky ReLU, Parametric ReLU) play an essential role in deep learning. The ReLU function is defined as:
\[
\text{ReLU}(x) = \max(0, x),
\]
and can be implemented in fixed-point arithmetic using a simple conditional statement. Leaky ReLU is defined as:
\[
\text{Leaky ReLU}(x) =
\begin{cases}
x, & \text{if } x > 0 \\
\alpha x, & \text{otherwise},
\end{cases}
\]
where $\alpha$ is a small constant. Parametric ReLU extends this concept by making $\alpha$ a learnable parameter. In our system, piecewise linear approximations are used to balance computational efficiency with the precision required over a wide range of input values.

\subsection{Loss Functions and Their Taylor Series Approximations}

Loss functions guide the training process by quantifying the error between predicted and true values. Common functions include the Mean Squared Error (MSE) and Cross-Entropy losses. Their Taylor series approximations allow for implementation in fixed-point arithmetic by replacing logarithmic operations with polynomial approximations. Table~\ref{tab:loss_taylor} summarizes these approximations.

\begin{table}[ht]
    \centering
    \caption{Taylor Series Approximations of Common Loss Functions}
    \label{tab:loss_taylor}
    \renewcommand{\arraystretch}{1.0} 
    \scriptsize
    \begin{tabular}{| >{\centering\arraybackslash}m{1.5cm} 
                    | >{\centering\arraybackslash}m{3.3cm} 
                    | >{\centering\arraybackslash}m{3.0cm} |}
        \hline
        \textbf{Loss Function} & \textbf{Mathematical Definition} & \textbf{Taylor Series Approximation (around 0)} \\
        \hline
        Mean Squared Error (MSE) & $L(y, \hat{y}) = (y - \hat{y})^2$ & $(y - \hat{y})^2$ \\
        \hline
        Binary Cross-Entropy & $L(y, \hat{y}) = -\left[y \log(\hat{y}) + (1-y) \log(1-\hat{y})\right]$ & $-y\left(\hat{y} - \frac{\hat{y}^2}{2} + \frac{\hat{y}^3}{3}\right) - (1-y)\left(-\hat{y} - \frac{\hat{y}^2}{2} - \frac{\hat{y}^3}{3}\right)$ \\
        \hline
        Categorical Cross-Entropy & $L(y, \hat{y}) = -\sum_i y_i \log(\hat{y}_i)$ & $-\sum_i y_i\left(\hat{y}_i - \frac{\hat{y}_i^2}{2} + \frac{\hat{y}_i^3}{3}\right)$ \\
        \hline
    \end{tabular}
\end{table}

\section{Results and Discussion}
\begin{figure}[ht]
\centering
\includegraphics[width=0.45\textwidth]{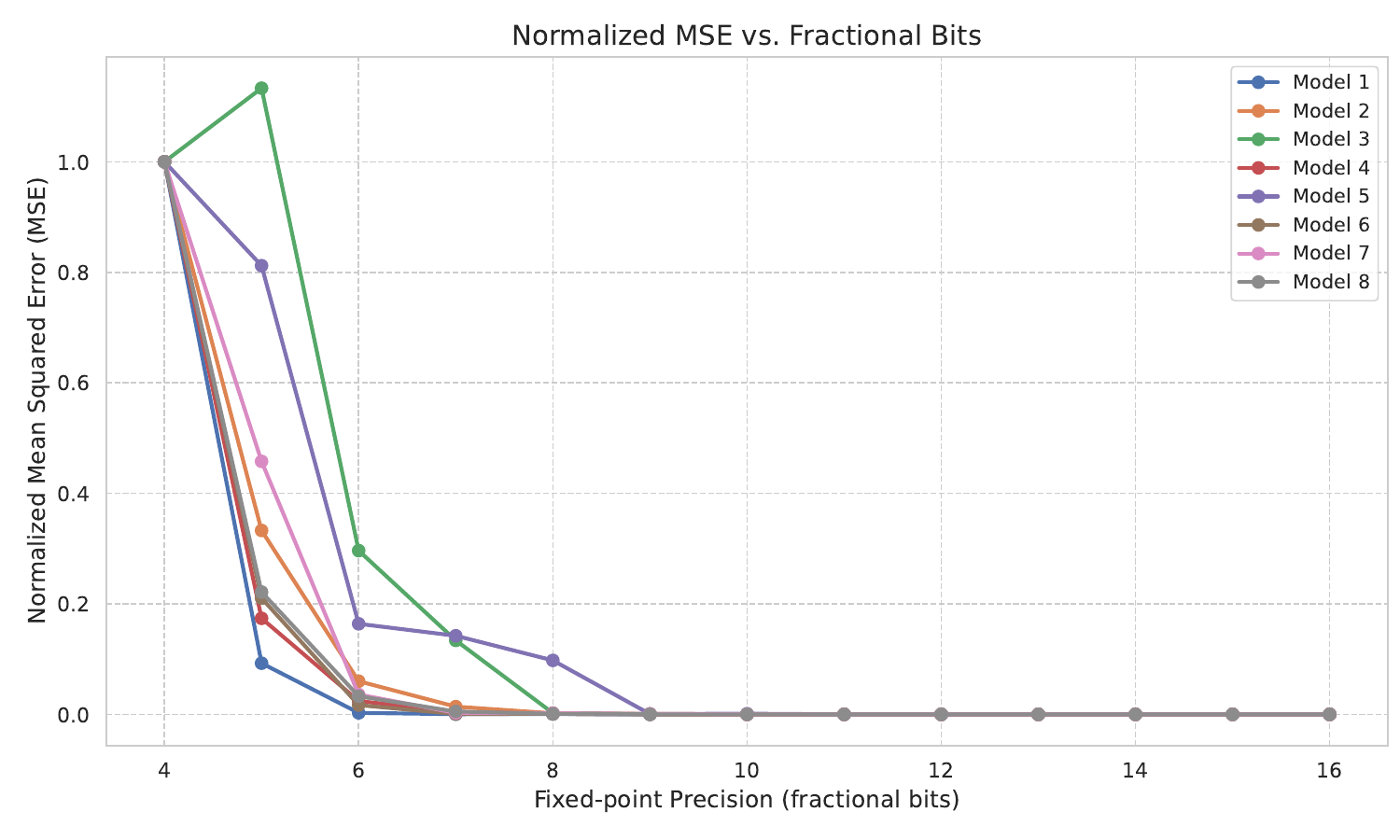}
\caption{
Normalized Mean Squared Error (MSE) versus fractional bit precision: Increasing fractional bits reduces quantization error but incurs overhead due to larger packet sizes, directly impacting network throughput. This highlights the trade-off between numerical precision and line-rate processing efficiency in P4-programmable FPGA SmartNICs.
}
\label{fig:mse_frac_bits}
\end{figure}

\begin{figure}[ht]
\centering
\includegraphics[width=0.45\textwidth]{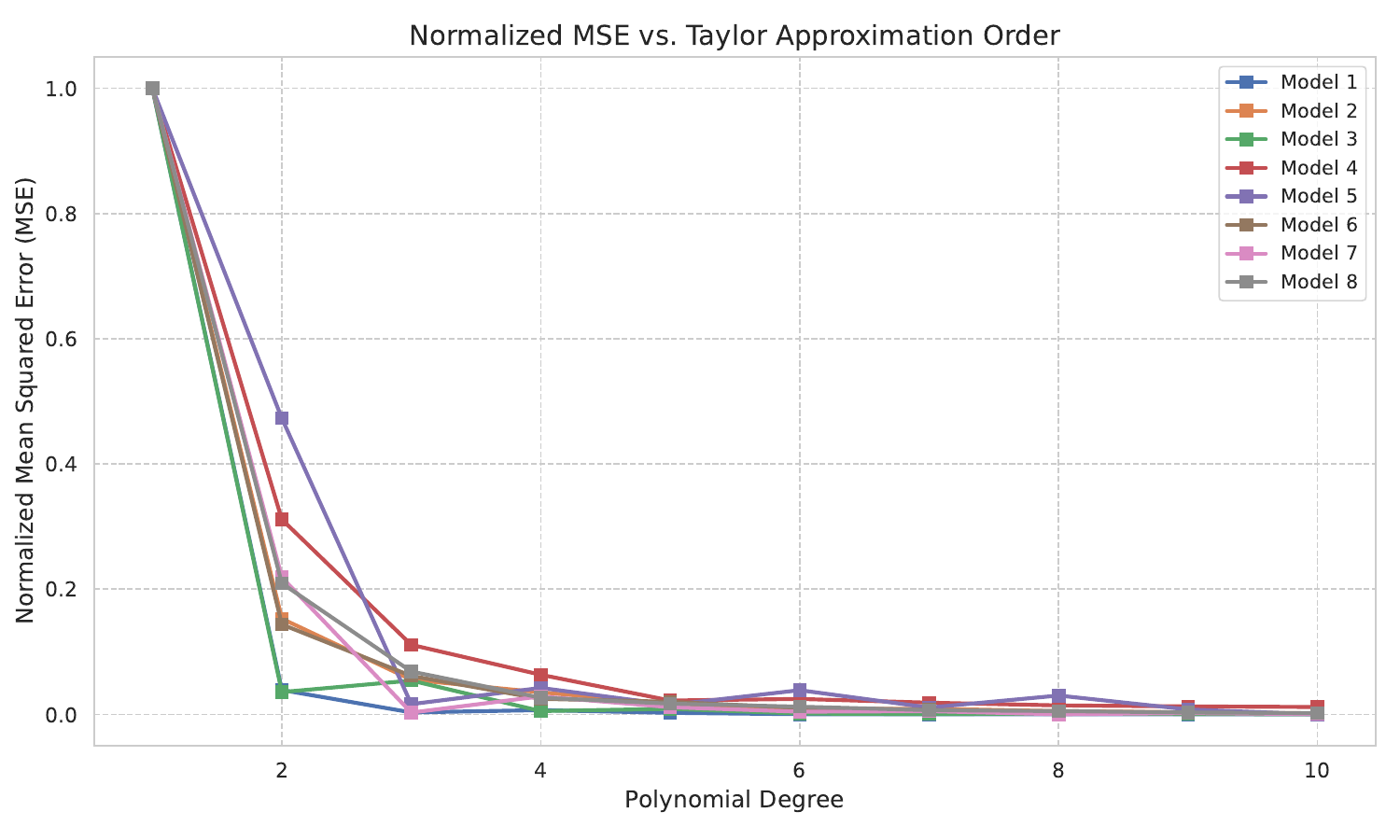}
\caption{
Normalized MSE versus Taylor polynomial order: Higher-order polynomials improve approximation accuracy for floating-point operations (e.g., sigmoid functions in neural networks) but require additional P4 table lookups and arithmetic stages, increasing pipeline latency and resource utilization.
}
\label{fig:mse_poly_order}
\end{figure}

Experiments demonstrate that in-network processing reduces inference latency to microsecond scale by eliminating PCIe round-trips. While fixed-point arithmetic and Taylor approximations introduce quantization errors, the normalized MSE remains below 0.15 for 8-bit fractional precision—a tolerable trade-off for latency-sensitive regression tasks like QoS prediction (Fig.~\ref{fig:mse_frac_bits}). Similarly, third-order Taylor polynomials balance accuracy and overhead (Fig.~\ref{fig:mse_poly_order}), limiting MSE to below 0.2 while requiring only two additional P4 table lookups per approximation. These results validate that lightweight ML models can operate at line rate without compromising network throughput. Future work will extend this methodology to support sampling for CPU training feedback loops to the control plane for continuous training on inference data, further bridging the gap between programmable data planes and edge-deployable models.

\section{Conclusion}
\label{sec:conclusion}
FPGA SmartNICs enable practical in-network machine learning without heavy compromise in throughput or burden in compute resources. The proposed framework supports diverse applications through a combination of fractional arithmetic and careful pipeline design. Future directions will focus on examining the performance of models with varying weight lengths, scales, degrees, and bit precision through an in-depth exploration of hyperparameter configurations.

\begin{acks}
This work was supported in part by National Science Foundation (NSF) awards CNS-1730158, ACI-1540112, ACI-1541349, OAC-1826967, OAC-2112167, CNS-2100237, and CNS-2120019. This paper was edited using LLMs hosted on the NRP~\cite{noauthor_nrp-managed_nodate}.
\end{acks}


\bibliographystyle{ACM-Reference-Format}
\bibliography{references}


\begin{thebibliography}{11}


\ifx \showCODEN    \undefined \def \showCODEN     #1{\unskip}     \fi
\ifx \showISBNx    \undefined \def \showISBNx     #1{\unskip}     \fi
\ifx \showISBNxiii \undefined \def \showISBNxiii  #1{\unskip}     \fi
\ifx \showISSN     \undefined \def \showISSN      #1{\unskip}     \fi
\ifx \showLCCN     \undefined \def \showLCCN      #1{\unskip}     \fi
\ifx \shownote     \undefined \def \shownote      #1{#1}          \fi
\ifx \showarticletitle \undefined \def \showarticletitle #1{#1}   \fi
\ifx \showURL      \undefined \def \showURL       {\relax}        \fi
\providecommand\bibfield[2]{#2}
\providecommand\bibinfo[2]{#2}
\providecommand\natexlab[1]{#1}
\providecommand\showeprint[2][]{arXiv:#2}

\bibitem[noa({[n.\,d.]})]%
        {noauthor_nrp-managed_nodate}
 \bibinfo{year}{[n.\,d.]}\natexlab{}.
\newblock \bibinfo{booktitle}{\emph{{NRP}-Managed {LLMs}}}.
\newblock
\urldef\tempurl%
\url{https://nrp.ai/documentation/userdocs/ai/llm-managed/}
\showURL{%
\tempurl}


\bibitem[bmv(2023)]%
        {bmv2_cite}
 \bibinfo{year}{2023}\natexlab{}.
\newblock \bibinfo{title}{BMv2: Behavioral Model v2}.
\newblock
\urldef\tempurl%
\url{https://github.com/p4lang/behavioral-model}
\showURL{%
\tempurl}
\newblock
\shownote{Online; accessed 2025-03-28}.


\bibitem[dpd(2023)]%
        {dpdk_cite}
 \bibinfo{year}{2023}\natexlab{}.
\newblock \bibinfo{title}{Data Plane Development Kit (DPDK)}.
\newblock
\urldef\tempurl%
\url{https://www.dpdk.org}
\showURL{%
\tempurl}
\newblock
\shownote{Online; accessed 2025-03-28}.


\bibitem[doc(2023)]%
        {docker_cite}
 \bibinfo{year}{2023}\natexlab{}.
\newblock \bibinfo{title}{Docker}.
\newblock
\urldef\tempurl%
\url{https://www.docker.com/}
\showURL{%
\tempurl}
\newblock
\shownote{Online; accessed 2025-03-28}.


\bibitem[esn(2023)]%
        {esnet_cite}
 \bibinfo{year}{2023}\natexlab{}.
\newblock \bibinfo{title}{ESnet SmartNIC}.
\newblock
\urldef\tempurl%
\url{https://www.github.com/esnet/esnet-smartnic-hw}
\showURL{%
\tempurl}
\newblock
\shownote{Online; accessed 2025-03-28}.


\bibitem[sca(2023)]%
        {scapy_cite}
 \bibinfo{year}{2023}\natexlab{}.
\newblock \bibinfo{title}{Scapy}.
\newblock
\urldef\tempurl%
\url{https://scapy.net/}
\showURL{%
\tempurl}
\newblock
\shownote{Online; accessed 2025-03-28}.


\bibitem[Bosshart et~al\mbox{.}(2014)]%
        {P4}
\bibfield{author}{\bibinfo{person}{Pat Bosshart}, \bibinfo{person}{Dan Daly}, \bibinfo{person}{Glen Gibb}, \bibinfo{person}{Martin Izzard}, \bibinfo{person}{Nick McKeown}, \bibinfo{person}{Jennifer Rexford}, \bibinfo{person}{Cole Schlesinger}, \bibinfo{person}{Dan Talayco}, \bibinfo{person}{Amin Vahdat}, \bibinfo{person}{George Varghese}, {and} \bibinfo{person}{David Walker}.} \bibinfo{year}{2014}\natexlab{}.
\newblock \showarticletitle{P4: Programming Protocol-Independent Packet Processors}.
\newblock \bibinfo{journal}{\emph{ACM SIGCOMM Computer Communication Review}} \bibinfo{volume}{44}, \bibinfo{number}{3} (\bibinfo{year}{2014}), \bibinfo{pages}{87--95}.
\newblock
\href{https://doi.org/10.1145/2656877.2656890}{doi:\nolinkurl{10.1145/2656877.2656890}}


\bibitem[Paolini et~al\mbox{.}(2021)]%
        {paolini2021programmable}
\bibfield{author}{\bibinfo{person}{Emilio Paolini}, \bibinfo{person}{Lorenzo De~Marinis}, \bibinfo{person}{Davide Scano}, {and} \bibinfo{person}{Francesco Paolucci}.} \bibinfo{year}{2021}\natexlab{}.
\newblock \showarticletitle{Programmable Switches for In-Networking Classification}. In \bibinfo{booktitle}{\emph{IEEE INFOCOM 2021 - IEEE Conference on Computer Communications}}. IEEE.
\newblock


\bibitem[Patel et~al\mbox{.}(2022)]%
        {patel}
\bibfield{author}{\bibinfo{person}{Shivam Patel}, \bibinfo{person}{Rigden Atsatsang}, \bibinfo{person}{Kenneth~M. Tichauer}, \bibinfo{person}{Michael H. L.~S. Wang}, \bibinfo{person}{James~B. Kowalkowski}, {and} \bibinfo{person}{Nik Sultana}.} \bibinfo{year}{2022}\natexlab{}.
\newblock \showarticletitle{In-network fractional calculations using {P4} for scientific computing workloads}. In \bibinfo{booktitle}{\emph{Proceedings of the 5th International Workshop on {P4} in Europe, EuroP4 2022, Rome, Italy, 9 December 2022}}, \bibfield{editor}{\bibinfo{person}{Marco Chiesa} {and} \bibinfo{person}{Shir~Landau Feibish}} (Eds.). \bibinfo{publisher}{{ACM}}, \bibinfo{pages}{33--38}.
\newblock
\href{https://doi.org/10.1145/3565475.3569083}{doi:\nolinkurl{10.1145/3565475.3569083}}


\bibitem[Sapio et~al\mbox{.}(2023)]%
        {sapio2023janus}
\bibfield{author}{\bibinfo{person}{Antonio Sapio}, \bibinfo{person}{Ibrahim Abdelaziz}, \bibinfo{person}{Abdulla Aldilaijan}, \bibinfo{person}{Marco Canini}, {and} \bibinfo{person}{Panos Kalnis}.} \bibinfo{year}{2023}\natexlab{}.
\newblock \showarticletitle{Janus: An Experimental Reconfigurable SmartNIC with P4 Programmability and SDN Isolation}. In \bibinfo{booktitle}{\emph{Proceedings of the 2023 ACM/SIGDA International Symposium on Field Programmable Gate Arrays}}. ACM.
\newblock


\bibitem[Smarr et~al\mbox{.}(2018)]%
        {smarr_pacific_2018}
\bibfield{author}{\bibinfo{person}{Larry Smarr}, \bibinfo{person}{Camille Crittenden}, \bibinfo{person}{Thomas {DeFanti}}, \bibinfo{person}{John Graham}, \bibinfo{person}{Dmitry Mishin}, \bibinfo{person}{Richard Moore}, \bibinfo{person}{Philip Papadopoulos}, {and} \bibinfo{person}{Frank Würthwein}.} \bibinfo{year}{2018}\natexlab{}.
\newblock \showarticletitle{The Pacific Research Platform: Making High-Speed Networking a Reality for the Scientist}. In \bibinfo{booktitle}{\emph{Proceedings of the Practice and Experience on Advanced Research Computing: Seamless Creativity}} (New York, {NY}, {USA}, 2018-07-22) \emph{(\bibinfo{series}{{PEARC} '18})}. \bibinfo{publisher}{Association for Computing Machinery}, \bibinfo{pages}{1--8}.
\newblock
\showISBNx{978-1-4503-6446-1}
\href{https://doi.org/10.1145/3219104.3219108}{doi:\nolinkurl{10.1145/3219104.3219108}}


\end{thebibliography}

\end{document}